
\input phyzzx
\input tables
\input epsf
\voffset = -0.4in
\footline={\ifnum\pageno=1 \nulline \else\newfootline \fi}
\def\nulline{{\hfill}}
\def\newfootline{\advance\pageno by -1\hss\tenrm\folio\hss}

\def\tbar{\bar{T}}
\def\ubar{\bar{U}}
\def\delgs{\delta_{\rm GS}}
\def\delgst{{\tilde{\delta}}_{\rm GS}}
\def\bpr{{ b'}}
\def\tt{\tilde T}
\def\gt{\tilde G}
\rightline {March 1994}
\rightline {QMW--TH--94/11}
\rightline {SUSX--TH--94/11}
\title {Anisotropic Solutions For  Orbifold Moduli\break
{}From Duality Invariant Gaugino Condensates.}
\author{D. Bailin$^{a}$, A. Love$^{b}$, W.A.
Sabra$^{b}$\ and \ S. Thomas$^{c}$}
\address {$^{a}$ School of Mathematical and Physical Sciences,
University of Sussex, Brighton BN1 9QH, U.K.}
\address {$^{b}$Department of Physics,\break
Royal Holloway and Bedford New College,\break
University of London,\break
Egham, Surrey, U.K.}
\address {$^{c}$
Department of Physics,\break
Queen Mary and Westfield College,\break
University of London,\break
Mile End Road, London,  U.K.}
\abstract {The values of the $T$ and $U$ moduli are studied for
those ${\bf Z}_N $ Coxeter orbifolds with the property that some of the
twisted sectors
have fixed planes for which the six-torus ${\bf T}^6 $ can not be
decomposed  into a direct sum  ${\bf T}^2\bigoplus {\bf T}^4 $ with the fixed
plane
lying in
${\bf T}^2 $. Such moduli in general transform under a subgroup of the modular
group $SL(2,Z).$ The moduli are determined by minimizing the effective
potential derived  from a duality invariant gaugino condensate. }
\endpage
\REF\one {L. J.
Dixon, V. S. Kaplunovsky and J. Louis,  {\it Nucl. Phys.}  {\bf B355} (1991)
649.}
\REF\two{J. P. Derendinger, S. Ferrara, C. Kounnas and F. Zwirner,
{\it Nucl. Phys.} {\bf B372 } (1992) 145. }
\REF\burt{ J. Louis, SLAC-PUB-5527 (1991); G. Lopes Cardoso and B. Ovrut,  {\it
Nucl. Phys. } {\bf B369 } (1992) 351, UPR-0481T (1991). }
\REF\three{I. Antoniadis, K. S. Narain and T. R. Taylor, {\it Phys.
Lett.}
{\bf B267} (1991) 37. }
\REF\nilles{ M. Dine, R. Rohm, N. Seiberg and E. Witten, {\it Phys. Lett. }
{\bf B156} (1985) 55; J. P. Derendinger, L. E. Ibanez and H. P. Nilles, {\it
Phys. Lett. } {\bf B155} (1985) 65; H. P. Niless, {\it Phys. Lett. } {\bf B115}
(1982) 193; S. Ferrara, L. Girardello and H. P. Nilles, {\it Phys. Lett. } {\bf
B125} (1983) 457;  C. Kounnas and M. Porrati, {\it Phys. Lett. } {\bf B191}
(1987) 91.}
\REF\four{L. E. Ibanez and D. Lust, {\it Nucl. Phys. } {\bf B382}
(1992) 305;
B. de Carlos, J. A. Casas and C. Munoz, CERN-TH6681/92 .}
\REF\five{  A. Font, L. E. Ibanez, D. Lust and F. Quevedo,
 {\it Phys. Lett. } {\bf B245} (1990) 401;
S. Ferrara, N. Magnoli, T. R. Taylor and G. Veneziano,
{\it Phys.Lett.} {\bf B245} (1990) 409;
M. Cvetic, A. Font, L. E. Ibanez, D. Lust and F. Quevedo,
{\it Nucl. Phys. } {\bf B361 } (1991) 194. }
\REF\di{L. Dixon, talk presented at the A. P. S.  D. P. F. Meeting at Housto
(1990); V. Kaplunovsky, talk presented at "Strings 90" workshop at College
Station (1990); L. Dixon, V. Kaplunovsky, J. Louis and M. Peskin, unpublished.}
\REF\lalak{ J. A. Casas, Z. Lalak, C. Munoz and G. G. Ross, {\it Nucl. Phys.}
{\bf  B347} (1990) 243.}
\REF\g{H. P. Nilles amd M. Olechowsky, {\it Phys. Lett. } {\bf B248}
(1990) 268; P. Binetruy and M. K. Gaillard, {\it Phys. Lett. } {\bf B253}
(1991) 119.}
\REF\j{ J. Louis, SLAC-PUB-5645 (1991).}
\REF\six{D. Lust and T. R. Taylor, {\it Phys. Lett. } {\bf B253}
(1991) 335.}
\REF\seven{D. Lust and C. Munoz, {\it Phys. Lett. }{\bf B279}
(1992) 272;
B. de Carlos, J. A. Casas and C. Munoz,
{\it Nucl. Phys. } {\bf B339 } (1993) 623.}
\REF\eight{P.  Mayr and S.  Stieberger, {\it Nucl. Phys.} {\bf B407} (1993)
725.}
\REF\nine{D. Bailin, A. Love, W. A. Sabra and S. Thomas,
{\it Mod. Phys. Lett.} {\bf  A9} (1994) 67.}
\REF\ten {D. Bailin, A. Love, W. A. Sabra and S. Thomas,
{\it Phys. Lett.} {\bf  B320} (1994) 21.}
\REF\eleven{B. de Carlos, J. A. Casas and C. Munoz,
{\it Phys. Lett.}{\bf B263} (1991) 248.}
The values of the untwisted  $T$ and $U$ moduli for an orbifold
are of importance because they enter the string loop threshold corrections
relevant to unification of gauge coupling constants [\one-\three] and the
soft supersymmetry
breaking terms derived from orbifold compactifications [\four].
The effective potential is flat in these moduli to all orders of string
perturbation theory
and some non-perturbative mechanism is therefore required to
provide an
effective potential that can determine the moduli. A promising
mechanism is
a non-perturbative superpotential due to
gaugino
condensation in the hidden sector [\nilles]. This  approach has been
applied to the
calculation of the moduli by incorporating target-space duality invariance
[\five-\seven]. The derivation of
this non-perturbative superpotential requires a knowledge of the moduli
dependence
of the string loop threshold corrections to gauge coupling
constants.
This moduli dependence is associated [\one] with orbifold
twisted sectors for which there is a complex plane of the six torus
${\bf T}^6$ fixed by the twists in both space and time worldsheet directions.
The threshold corrections take
the standard
form [\one-\three ] with $SL(2,Z)$ modular symmetry group when
for all such
fixed planes, ${\bf T}^6$ can be decomposed into a direct sum ${\bf T}^2
\bigoplus
{\bf T}^4 $
with the fixed plane lying in ${\bf T}^2$. The moduli dependent
threshold
corrections have also been calculated for all ${\bf Z}_N$ Coxeter
orbifolds for which this simplifying assumption does not apply [\eight,
\nine].
We shall refer to these as non-decomposable orbifolds.
For non-decomposable orbifolds the threshold corrections  have modified forms
with
modular symmetry groups [\eight-\ten ] which are subgroups of $SL(2,Z)$,
not always
the full group. It is our purpose here, to study the values of
the moduli for such non-decomposable  ${\bf Z}_N$ orbifolds.
The lattices for which ${\bf Z}_N$ point groups provide such orbifolds are
listed in Table $1$.

Previous work [\five-\seven] on minimizing the effective
potential to determine the moduli has been confined, for convenience,
to the isotropic case where the three generic moduli
$T_i $, $i = 1,2,3$, have been
set all equal so that there is a single overall moduli $T$.
For the
non-decomposable orbifolds discussed, this is no longer
an appropriate assumption because the string loop threshold
corrections place $T_1$, $T_2 $ and $T_3 $ on a different footing.
We shall also include the contribution of the $U$-moduli
to the effective potential which has previously been neglected. However,
we shall confine our attention to the pure gauge case without matter
fields coupling
to the hidden sector gauge group. Since we are dealing with
modular symmetry
groups that are subgroups of $SL(2,Z)$, duality anomaly cancellation
considerations
continue to lead to a Kahler potential $K$ of the form
$$ K \,=\, - {\rm ln }\, y - \sum_i \, {\rm ln}\, (T_i + \tbar_i ) -
\,\sum_m{\rm ln}\,
  (U_m +\ubar_m)\eqn\un$$
where
$$ y\, =\, S + \bar{S} - {\sum_i}\, {{\delgs^i}\over{8 \pi^2}}\, {\rm
ln}\, (T_i + \tbar_i )
- \sum_m \, {{\delgst^m }\over{8 \pi^2}}\, {\rm ln}\, (U_m + \ubar_m).
\eqn\deux$$
In \un\ and \deux, $S$ is the dilaton field, the sum over $i$
runs from 1 to 3 and the sum over $m$ from 1 to the hodge number $h_{(2,1)} $.
Off-diagonal moduli, $T_{ij}$ which are present for some ${\bf Z}_N$ orbifolds,
but never
contribute to string loop threshold corrections have been dropped.
The Green-Schwarz coefficients $\delgs^i $ and $ \delgst^m $ are required
[\two, \burt] for the
cancelation of the anomalies of the underlying $\sigma$-model. In this case, in
order for the Kahler potential to transform as at tree level under modular
transformations, one should allow a non-trivial modular transformation of the
dilaton field $S.$
In what follows it
will  usually
be convenient for notational purposes to treat the $U$ moduli as
additional
$T$ moduli rather than to display them separately. In general,
the effective
potential $V$ is given by [\five-\seven]
$$\eqalign{ y \prod_i \, (T_i + \tbar_i )\, V\, = &  \,
\vert W - y \, {{\partial W}\over{\partial S}}
\,{ \vert }^2  - 3 \, \vert W {\vert }^2 \cr
 +& \sum_i {{y}\over{(y -{\textstyle{\delgs^i}\over{\textstyle 8
\textstyle\pi^2}}) } }\vert W -
{ {\delgs^i}\over{8 \pi^2} }{ {\partial W}\over{\partial S}} - (T_i +
\tbar_i )
{{\partial W}\over{\partial T_i } }{ \vert }^2  } \eqn\trois $$
 where $W$ is the superpotential and the sums and products over
$i$ are now
 understood to include the $U$ moduli.

To be able to fix the values of the moduli, the modular
invariant gaugino condensate superpotential is required. This in turn, requires
a knowledge of the string loop threshold
corrections to the
gauge coupling constants.
In general, for the factor $G_a $ of the hidden sector gauge
group,
 the running coupling constant $g_a(\mu)$ obeys
 $$g_a^{-2}(\mu ) \, = \,  g_a^{-2} (M )   + {{b_a}\over{16
\pi^2 }}\,
 {\rm ln}\, ({{M^2}\over{\mu}^2} ) + {{\triangle_a }\over {16 \pi^2}},
\eqn\quatre$$
where $M$ is the string scale and $\triangle_a $ is the threshold
correction.
 For the non-decomposable ${\bf Z}_N$ orbifolds (in the sense discussed
above),
 $\triangle_a $ takes the general form  [\eight , \nine ]
 $$\triangle_a \, = \,-\sum_i ( \bpr^i_a - \delgs^i )\, \Big( {\rm ln }\, (T_i
+ \tbar_i )
+ \sum_m \, {{C_{im}}\over{2}}\,  {\rm ln }\, \vert \eta \,
({ {T_i}\over{l_{im}} } )\,  {\vert
 }^4 \Big) \eqn\cinq$$
 where the range over which $m$ runs depends on the value of $i$,
 $$\sum_m C_{im} = 2 \eqn\sixx$$
 for all $i$. The coefficients $C_{im}$ and  $l_{im} $ are
 given in Table 2, for the various non-decomposable Coxeter orbifolds.
 The $\bpr^i_a $ are the usual coefficients
[\two, \burt] which
 are determined by the contribution of the massless states to the
duality
 anomaly in a way which does not depend on the underlying lattice
for the
 orbifold,
 $$\eqalign{ \bpr^i_a \, =& \, - C(G_a ) + \sum_{R_a}\,  T(R_a )\, (1 + 2
n^i_{R_a} ), \qquad \hbox {for} \   T\  \hbox {moduli}\cr
\bpr^i_a \, =& \, - C(G_a ) + \sum_{R_a}\,  T(R_a )\, (1 + 2 l^i_{R_a} ),
\qquad  \hbox {for} \  U\  \hbox {moduli}.}\eqn\sept$$
 In \sept, $C(G_a )$ and $T(R_a )$ are Casimirs for the gauge
group factor
 $G_a $ and its representations  $R_a $, $n^i_{R_a }$ and $l^i_{R_a }$, $i =1,
2, 3 $  are the modular weights under $T$ and $U$ duality transformation
respectively, for the massless states in the
representation
 $R_a $ of $G_a $. For the pure gauge case, where there is no
matter charged
 under the hidden sector gauge group, (7) simplifies to
 $$\bpr^i_a \, = \, -C(G_a )={{b_a}\over {3}}.\eqn\huit $$
 The holomorphic part $f^{W}_a $ of the gauge kinetic function is
given by
 $$f^W_a \, =\,  S - \sum_{i,m}\, { {(\bpr^i_a - \delgs^i )}\over{8
\pi^2 } }\,
 C_{im}\,  {\rm ln }\, \Big(\eta ({ {T_i}\over{l_{im}} } ) \Big) \eqn\neuf $$
 with $S$ the dilaton field whose real part $S_R$ controls the
value of the
 gauge coupling constant at the string unification scale \footnote*{
  Strictly speaking, when $\delgs^i $  and $\delgst^m  \, \neq \, 0 $,
it is the $y$ field given in \deux\ which  takes on this role.}

 The modular invariant superpotential $W_{np}$ is of the form
[\five, \six, \eleven, \lalak]
 $$ W_{np}\, =\, \sum_a \,  {W_{np}}^a  \,= \, \sum_a \,  d_a \, e^{{ {24 \pi^2
}\over{b_a}} f^W_a }
  \eqn\dix$$
and \neuf\ gives
$$ {W_{np}}^a \,= \,  d_a \, e^{{ {24 \pi^2}\over {b_a}} S}\, \prod_{i,m}
{\Big(\eta ({ {T_i}\over{\l_{im}} } ) \Big)}^{{-C_{im}}\,  (1 - { {3
\delgs^i}\over{b_a}})}\eqn\onze $$
Notice that \sixx, together with the non-trivial modular
transformation of the dilaton field $S$ beyond the tree level [\two, \burt],
ensures
that $W_{np}^a$ has the correct modular weight of $-1$ for modular
transformations on each of the $T_i $ separately.

 When there is at least one complex plane that is rotated by all
non-trivial powers of the point group generator $\theta ,$ $i.e$., at least one
$N=1$ complex plane, then the fact that such a plane does not contribute to the
moduli dependent
 threshold corrections implies, [\seven ]
  $$ \bpr^i_a \, = \, {{b_a}\over{3}} = \delgs^i \eqn\douze $$
  so that $b_a $ is independent of $a$,
$$ b_a \, =\, b, \quad\quad\quad {\rm for}\,{\rm  all }\,  a. \eqn\treize $$
In these circumstances, each of the $W_{np}^a $ has the same
  $e^{24 \pi^2 S/ b} $ dependence  on the dilaton
field and
  there is effectively a single gaugino condensate. When $W =
W_{np} $ is substituted in \trois, this results in there being no
minimum for $S_R$ and the value of the dilaton field not being
determined.
 We shall assume with previous authors [\five , \six ] that some
unspecified mechanism fixes the dilaton field at a value in the
region of weak coupling. Following [ \seven ] ,
 we shall take
 $S_R\, \approx \, 2$.
 (One mechanism that has been discussed in the
literature is for there to be charged matter [\lalak, \seven, \eleven] coupled
to the hidden sector
\REF\ta{ N. V. Krasnikov, {\it Phys. Lett. } {\bf B193}
(1987) 37; T. R. Taylor, {\it Phys. Lett. } {\bf B252}
(1990) 59.}
gauge group with more than one condensate [\di, \lalak, \ta].
An intermediate scale mass for this matter can be generated by
the expectation values of certain singlet scalars, which can be `integrated
out' by
  minimizing the effective potential. It can then sometimes
happen that
  there is a realistic minimum for the dilaton expectation value
and that  the
  $T_i $ dependence of the effective potential is the same as for
the pure gauge case at least for $\delgs^i = 0.$
Similar calculations can be carried out here, with the same
outcome.
  However, the particular mechanism for fixing $S_R$ can not work
for ${\bf Z}_N$ orbifolds
  because there is always at least one $N=1$ complex plane and,
consequently,
  effectively a single gaugino condensate, even in the presence
of matter, because of the requirement of correct modular weight for the
gaugino condensate superpotential.)

Once the dilaton expectation value $S_R $ has been fixed at a
value in the region of weak coupling, the effective potential \trois\ with
$W  = W_{np} $ given by \dix\ and \onze\ may be minimized to find the
values of the moduli $T_i $ (including the $U$ moduli.)
 The $N= 1$ moduli will be exceptions to this procedure, since they do not
enter the threshold correction formula \cinq\ . We will discuss their role
more fully later.
Because of the invariance of the effective potential
under modular transformations on each modulus separately there will,
generally speaking,
be extrema at the fixed points of modular transformations (the self-dual
points.)
For the case of an $SL(2,Z)$ transformation
$$T_i \, \rightarrow \,  {{a_i T_i  - i b_i }\over{i c_i T_i  + d_i } }
  \eqn\quatorze $$
where $a_i , b_i , c_i $ and $d_i $ are integers with $ a _i d_i - b_i c_i  =
1. $ The fixed points [\five ] are at
$\tt_i = i$ or $e^{\pi i/3}$ where
$\tt_i=i T_i $
and they are fixed by the modular transformations ${\cal S} $ and
${\cal T}{\cal S} $, respectively, where
$$\eqalign{ {\cal S}:& \qquad \tt \rightarrow - {1\over\tt},\cr
{\cal T}: & \qquad \tt \rightarrow \tt + 1.}
\eqn\ntn $$
Here, for non-decomposable ${\bf Z}_N $ orbifolds, the relevant modular
symmetries for the $T$ and $U$ moduli [\eight-\ten] deriving
from the
string loop threshold corrections are in general congruence
subgroups
of $SL(2,Z)$. The congruence subgroups that arise are $\Gamma_0 (2)$,
$\Gamma_0 (3)$ , $\Gamma^0 (2) $ and $\Gamma^0 (3) $ where
$\Gamma_0 (n) $
and $\Gamma^0 (n) $ are represented by the matrices
$$ \Gamma_0 (n) \qquad : \pmatrix{a&b \cr c &d } , \qquad \qquad
ad - bc = 1,\quad
c = 0\,  ({\rm mod } \ n) \eqn\twnty $$
and
$$\Gamma^0 (n) \qquad : \pmatrix{a&b \cr c &d } , \qquad \qquad
ad - bc = 1, \quad
b = 0\,  ({\rm mod } \ n)\eqn\tnyo $$
The corresponding fundamental domains are
$$\eqalign{\Gamma_0(2): \qquad {\cal F}_1 =& \{ {\cal I}, {\cal
S}, {\cal S} {\cal T} \}
{\cal F},\cr
\Gamma_0(3): \qquad {\cal F}_2 =& \{ {\cal I}, {\cal S}, {\cal S}
{\cal T} ,
{\cal S}{\cal T}^2 \} {\cal F},\cr
\Gamma^0(2): \qquad {\cal F}_3 =& \{ {\cal I}, {\cal S}, {\cal T}
{\cal S} \}
{\cal F},\cr
\Gamma^0(3): \qquad {\cal F}_4 =& \{ {\cal I}, {\cal S}, {\cal T}
{\cal S},
{\cal T}^2 {\cal S} \} {\cal F}}\eqn\tntt $$
where
$$ {\cal F} = \{\tt : {\rm Im } \tt > 0, \quad \vert {\rm Re
}\tt \vert <
{1\over 2} , \quad \vert \tt \vert > 1 \} \eqn\tnth $$
is the usual fundamental domain for $SL(2,Z)$. The fixed points for
the congruence
subgroups can now be found from the fixed points of $SL(2,Z)$ by
noticing that
if $z$ is  a fixed point of the transformation $L$ lying in
$\alpha
{\cal F} $, with $\alpha = {\cal S}, {\cal S} {\cal T}, {\cal
T}{\cal S},
{\cal S}{\cal T}^2$ and $ {\cal T}^2 {\cal S} $, then ${\alpha
}^{-1} z $
is a fixed point of ${\alpha}^{-1} L  {\alpha}  $ lying in ${\cal
F} $,
with the outcome that the possible fixed points are
$T_i ={\textstyle{1 + i}\over\textstyle{2}},$ $T_i =1 - i$  for  $\Gamma_0 (2)$
and
$\Gamma^0 (2)$ respectively, $T_i ={\textstyle{1 + i \sqrt{3}
}\over\textstyle{2 \sqrt{3}}}$
for $\Gamma_0 (3)$ and $T_i  ={\textstyle\sqrt{3}\over\textstyle 2} ( 1 - i
\sqrt{3} ) $ for $\Gamma^0 (3).$  These points are fixed by the elements ${\cal
S}{\cal T}{\cal S}({\cal S}{\cal T})^{-1}\in \Gamma_0(2),$ ${\cal T}{\cal
S}{\cal T}^{-1}\in \Gamma^0(2),$  ${\cal S}{\cal T}^2{\cal S}({\cal S}{\cal
T})^{-1}\in \Gamma_0(3)$ and  ${\cal T}^2{\cal S}{\cal T}^{-1}\in \Gamma^0(3).$
In  the
$SL(2,Z)$ case with a single overall modulus [\five],  and
$\delgs^i \, = \, 0 $,
the extrema associated with the two fixed points in the fundamental
domain ${\cal F} $,
$T = 1,\,  e^{ -\pi i/6}\,$ are a saddle point
and a maximum respectively, with the minimum  lying close to them.
However as noted in [\five , \six ] , the exact position of the minima in $T$
relative to the fixed points depends on the value of $S_R $ .
 For the case of the subgroups,
we see that there is a single fixed point in the corresponding
fundamental region.  The connection between this
fixed point and extrema of $V$ is rather more involved
since unlike [\five ],  we do not assume that all the $\delgs^i $ are
necessarily vanishing. We will return to this issue later.

To carry out the minimization of the effective  potential \trois\ to determine
the true minima, values of $\delgs^i $ are required. These may be
determined
for any particular orbifold by a comparison of the expression
given
in ref. [\two] for the string loop threshold corrections in terms of
$\bpr^i_a $ and $\delgs^i $
 with the expression given in [\one] in terms of the
renormalization
 group equation coefficients ${(b^{{N}=2}_a )}^i $ for the $N=2$
 orbifold ${\textstyle {{\bf T}^6}/\textstyle{G_i }} $ where $G_i $ is the
subgroup of
the point group
 $G$ that leaves the $i$-th complex plane unrotated. Thus,
 $$\delgs^i \, = \, \bpr^i_a -    {(b^{{N}=2}_a )}^i \, {{\vert G_i
\vert }\over{\vert G \vert}}. \eqn\tfiv$$
 This expression applies equally for $T$ and $U$ moduli provided
we use the appropriate modular weights [\four ] in the calculation of
$\bpr^i_a $.
 In the case of a pure gauge hidden sector
 $$\delgs^i \, =\,  {{b_a}\over 3}\,  ( 1 - 2 \, { {\vert G_i \vert }\over
{\vert G \vert}}). \eqn\tsix$$
For the ${\bf Z}_N $ orbifolds considered here (and with
$b \, = \, -90 \, \, $for $E_8 $ )
$$\eqalign{\delgs^3 =0, \qquad & Z_4, \quad Z_8 - II \cr
\delgs^1 = -10, \qquad \delgs^3 =0 , \qquad &Z_6 - II \cr
\delgs^3 =-10, \qquad &Z_{12} - I. }\eqn\twsev $$
For moduli associated with $N=1$ planes, $\delgs^i \,  $ is given by
\douze. Whilst the task of minimizing the potential \trois\ in the $T$ and
$U$ moduli is best carried out numerically, it is instructive to calculate
$\partial_{T_i}V$ and $\partial_{S} V$, in order to see the connection between
fixed points of the modular
(sub)groups discussed above, and certain extrema of $V$. One finds that
the extrema conditions for the $T_i $ and $S$ moduli are
$$\eqalign{({{3\delgs^i}\over{b}} -& x ) \, \sum_{j} \,
{{x}\over{( x - {\textstyle{3\delgs^j}\over\textstyle{b}} )}} \vert \gt^j
{\vert }^2
+ x( T_i + \tbar_i )  \, {3\delgs^i}\sum_{j} \, \{
{{{3\delgs^j}\over{(b x - 3 \delgs^j)}^2 }} \, {{\vert \gt^j {\vert}^2 }\over
{(T_i + \tbar_i ) } } \cr
 + & {{x}\over{( x - {\textstyle{3\delgs^j}\over\textstyle{b} } ) }} [ (
{{W_{T_i}}\over{W}}
 (1- {{3\delgs^j}\over{b} } ) - {\delta}_{ij}  {{W_{T_j}}\over{W}}-
 (T_j + \tbar_j ) {W_{T_i T_j }\over W} ) \, {\bar{ \gt}}^j \cr
 -& {\gt}^j\, ( {{\bar{W_{T_j}} }\over{\bar{W}} }) {\delta}_{ij} ] \, \} -
 x\, {\gt}^i \, (x^2 - 2x -2 ) + {{3\delgs^i}\over{b} }(1-x)(x^2 - 2)
\, = \, 0  }\eqn\tweig $$
and
$$ (2- x^2 )(1-x) + \sum_i \,    {({{x}\over{( x -
{\textstyle{3\delgs^i}\over\textstyle{b} } )}})}^2 \, [ \, \vert \gt^i
{\vert}^2 \,
(x -  {{3\delgs^i}\over{b}}-1 )\, ]  \, = \, 0 \eqn\twenin$$
where in \tweig\ and \twenin\ , $ x =  ({24 \pi^2} /b )\, y $ and  the
functions $\gt^j $ are defined
as
$$ \gt^j \, = \, (1 - {{3\delgs^j}\over{b}} ) (T_j + \tbar_j ) \, {\hat{G}}^j
\eqn\thiry $$
with
$$ {\hat{G}}^j \, = \, ( {1\over{T_j + \tbar_j } } + \sum_m \, C_{jm} \,
{{\partial_{T_j} {\eta}({ {T_j}\over{l_{jm}} } ) }
\over{{\eta}({ {T_j}\over{l_{jm}} } ) }} \, )\, = \,
{1\over{T_j + \tbar_j } } + G^j_2  \eqn\thron$$
In  \thron, $G^j_2 $ are weight 2 Eisenstein functions for the corresponding
modular subgroups discussed above. Since
${\hat{G}}^j \, \sim  ({\partial \triangle } )/ ( \partial T_j )$, where
$\triangle$ is the
threshold correction \cinq\ for a simple gauge group, it is clear that
${\hat{G}}^j $ vanish at the fixed points of the corresponding
 modular subgroups. (${\hat{G}}^j $
are the generalization to subgroups of $SL(2, Z)$, of the  weight 2
modular functions  transforming under the full $SL(2, Z) $ group described
in [\five].) It is also clear from \tweig\ that such fixed points will be
extrema of  $V$ provided $\delgs^j \, = \, 0 $ and (or) $x $ is  restricted
to its extremum value. We could have arrived at the same conclusion
by symmetry considerations alone. If $V(S, T) $ is invariant
under a modular (sub)group in which $T_i $ transform as in \quatorze\
then recalling that for $\delgs^i \neq 0 $, $S$ itself transforms as
$$ S \, \rightarrow \, S - \sum_i \, { {{\delgs}^i}\over{8 \pi^2 } } \, {\rm ln
}\,
( i c_i T_i + d_i) \, = S^{' } \eqn\thrt $$
one has
$$ {{\partial V(S,T)}\over{\partial T_i}} \, = \,  {1\over{( i c_i T_i + d_i
)}^2 }
\, {{\partial V(S^{'}, T^{'} )}\over{\partial T^{'}_i } } -
{{i c_i {\delgs^i} }\over{8 \pi^2 (i c_i T_i + d_i ) } } {{\partial V(S^{'},
T^{'}_i  ) }\over {\partial S^{'} } } \eqn\thrthr $$
Using the fact that $\vert i c_i T_i  + d_i \vert \, = \, 1 , \,
i c_i T_i  + d_i \neq \pm 1 \, $ at fixed points, we arrive at the
above conditions for such points to be extrema. (Note that for fixed
points, $S^{'}  \, = \, S $ up to axionic shifts, which leave $V$ invariant.)

As stated earlier, the position we adopt in this paper is to fix
the value of $S_R \, \approx \, 2$ , which will not be
an extremum of the potential  $V$ ( at least not in the pure gauge
case.) As a result, the connection between fixed points of
 modular subgroups and extrema of $V$ will not exist
for those moduli having non-vanishing Green-Schwarz
coefficients.
We will now consider a specific example which will make these points
clear and
illustrate the
differences between effective potentials that are invariant under modular
subgroups of $SL(2, Z)$ rather than the full $SL(2,Z)$ as
considered in [\five]. An interesting example is that of
the $Z_6-II-b $ Coxeter orbifold. There are $3$  moduli $T_i  , \,
i = 1,2,3 \, $,
and a modulus $U_3 $ which enter the effective potential, where the subscript
refers to the
corresponding complex plane. As
demonstrated in [\nine, \ten], the corresponding threshold corrections
are invariant under ${\Gamma}^{0}_{T_3} (3) , \,
{\Gamma}^{0}_{U_3  -2 i }(3) , \,$and $ {SL}_{T_1} (2,Z) \, $ respectively,
whilst $T_2 $ being a modulus associated with an $N=1$ plane,
does not contribute to  the threshold corrections.
The symmetry of the threshold
correction transfers itself to that of $V$, even for $U_3 - 2 i $
because  the Kahler potential only depends on Re $(U_3 )$.
 However, note that the
symmetry of $V$ with respect $U_3 $ is not
$\Gamma^{0} (3) $ but a different subgroup. This subgroup was
identified in [\nine ] as being
$$ {U_3} \, \rightarrow \, { { { a} \,
{U_3} -i { b }}\over{i { c }  {U_3} +{ d }} }, \qquad
( d + c ) - (a + b ) \, = \, 0 \, {\rm mod}\, 3
\eqn\thrfo $$

  Since $\delgs^2 = -30 $  (see \douze\ )
$T_2 $ does not have a flat potential.  Nevertheless, the essential role of
this field is
 to shift the value of $S$
by terms logarithmic in $T_2 $.  Since it is the field $y$  (defined in
\deux\ ) which controls the value of the gauge coupling
at the string scale, and not just $S_R $, whatever mechanism
fixes the latter at realistic values must also give realistic values
to $T_2 $. For now, we will set $T_2 \approx 1 $.
Henceforth, we  concentrate on the moduli
 $T_1, T_3 $ and $ U_3 $ which have rather more non-trivial
potentials. We find that
a numerical minimization of $V$ yields the values
$$ T_1 \,  = \, 1.39 + i(0.15 + n)\, ,\quad T_3 \, = \, 0.87 + i (1.50 + 3m)\,
, \, \quad
U_3 \, = \, 0.87 + i (3.50+ 3 p )         \eqn\thrfiv$$
where in \thrfiv \   $n,m $ and $p$ are arbitrary integers. In
figures 1-3,  $V$ is plotted in the 3 moduli respectively; in
each case two of the three   moduli are evaluated at the minimum values given
above.  Fig.1\footnote*{ In this and  subsequent plots,
the potential $V $ has been rescaled by a factor of \break
$-({24 \pi^2}/b)e^{48\pi^2S_R/b}$}shows the dependence of $V$ in $T_3 $,
and shows the characteristic periodicity in  Im( $T_3 $) except
that unlike the $SL(2,Z)$  case, ( as for example
occurs for the modulus $T_1 $ shown in Fig.2),
 the periodicity is  $3m$ rather than
$m$. This is due to the fact
that  the axionic shifts in $\Gamma^{0} (3) $ are generated by  ${\cal T}^3$
and not ${\cal T}$.  The
minima shown occur at the  fixed points of $\Gamma^{0} (3) $.
This is a consequence of taking $S_R \, \approx \, 2$. For larger
values of $S_R $ (the exact value depending on the
particular orbifold in question),
 one finds that the  minima do not occur at the fixed points,
the latter now being  associated with saddle points. This behavior
for large $S_R$ is easy to verify by evaluating the matrix
of second derivatives of $V$ at the fixed points. The result  is
$$\eqalign{ V_{T_i T_j }{\vert}_{f.p.} \, = &\,  \rho_i \, [ \, \delta_{ij} \,
( \,
\gt^{i,i} \, x(2-x) + {1\over{T_i + \tbar_i } } \, ) - {1\over{T_j + \tbar_j }
} \, ] \cr
V_{T_i \tbar^j}{\vert}_{f.p.}\, = & \, \rho_i \, [ \, \delta_{ij} \, ( \,
(T_i + \tbar_i ) \vert \gt^{i,i} {\vert}^2  - \gt^{i,i} - {{1 + 2x - x^2
}\over{
T_i + \tbar_i }  } \, ) + \gt^{i,i} \, {{T_i + \tbar_i }\over{T_j + \tbar_j } }
\, ] }\eqn\new$$
where in \new\ , $\gt^{i,i}\, = {\partial \gt^i} / {\partial T_i } \, $
and $\rho_i \, = \,  - \vert W {\vert}^2 \, ( x \prod_j (T_j + \tbar_j ) \,
(T_i + \tbar _i ) )^{-1} \, $. Here we have only included those moduli for
which $\delgs^i $ is vanishing, since as we have seen previously,
it is only for these moduli that a simple connection exists
between fixed points and extrema. The eigenvalues of the matrix
in \new\ are $\lambda_i \, = \pm \, (x^2 - 2x -1 )\, (T_i + \tbar_i )^{-1}
\, \rho_i \, $ in the large $S_R $ limit, so fixed points are
 saddle points of $V$.

Although the values of the moduli at the minimum of $V$
do depend on the value of $S_R$, they  tend to a fairly
constant  limit for  $S_R \geq  6 $ , given by $ T_1 \,  = \, 1.35 + i(0.43 +
n)\, ,\quad T_3 \, = \, 1.57 + i (1.27 + 3m)\, , \, \quad
U_3 \, = \, 1.62 + i (3.33+ 3 p ) $ .

 In fig. 2 we see a much `flatter' potential for $T_1 $ than for
$T_3 $ and $U_3 $. Furthermore  the fixed points
of $SL(2,Z) $ do not correspond to extrema in this plot.
This  is  a consequence of
$\delgs^1 \, \neq \, 0 $ and the fact that we  have to choose $S_R $ away from
its extremum.
Indeed it should be mentioned that generally speaking,
  those moduli $T_i $ and
$U$ possessing non-vanishing Green-Schwarz coefficients
$\delgs^i $ and $\delgst^m $, will have their corresponding
modular symmetries broken by giving  $S$ an
expectation value that  does not
correspond to an extremum. (In the present example, this
affects only the $T_1$ modulus.)

Finally, fig. 3  shows the dependence on the $U_3 $ modulus which has
a similar appearance to the plot in  Fig. 1 , except that there is a relative
shift by a factor of 2 in Im($U_3 $).  Again the periodicity of
$3p$ can be understood from the fact that the subgroup of
$SL(2,Z) $ acting on  $U_3 $ (defined in \thrfo\ ), contains axionic
shifts generated by  ${\cal T}^3 $ rather than ${\cal T} $. Once
more  the
values of $U_3 - 2i $ at the minima   are  also
fixed points of $\Gamma^{0} (3) $.

To conclude, we have investigated the mechanism for  fixing
some of the moduli in orbifold compactifications due to
duality invariant gaugino condensation, where the duality
groups are not $SL(2,Z) $ but subgroups
which are different for different moduli. We have seen that
this  leads naturally to  anisotropic values for the moduli, in
contrast to the isotropic solutions previously found
 in the literature. The values of those  moduli
for which $\delgs^i \, = \, 0 $ always
occur at the fixed points of the duality group for
$S_R  \approx 2.$  \vskip 2cm
\centerline {\bf{ACKNOWLEDGEMENT}}
This work is supported in part by S.E.R.C. and the work of S.
Thomas is
supported by the Royal Society.
\vskip2cm
\centerline{\bf {Table Captions}}
\noindent
{\bf Table}. 1. Non-decomposable $Z_N$ orbifolds. For the point
group generator $\theta $, we display $ ( \zeta_1 , \zeta_2 , \zeta_3 ) $
such that the action of $\theta $ in the complex plane  orthogonal
basis is $(e ^{2\pi i \zeta_1}, e ^{2\pi i \zeta_2},e ^{2\pi i \zeta_3} ). $
\vskip0.5cm
\noindent
{\bf Table}. 2. Values of $C_{im} $ and $l_{im} $ appearing in the threshold
corrections for non-decomposable orbifolds.
\vskip0.5cm
\centerline{\bf{Figure Captions}}
\noindent
{\bf Fig}. 1. The dependence of $V $ on the $T_3 $ modulus for the
orbifold  $Z_6-II-b $. Minima occur at the fixed points of the
subgroup $\Gamma^{0}(3) \, $, $ T_3 \, = \, \sqrt{3} /2 + i\,(1.5
+ 3 m) $.
\vskip0.5cm
\noindent
{\bf Fig}. 2. The dependence of $V$ on the modulus $T_1$ for
the orbifold $Z_6 - II-b $. Minima occur at $T_1 \,  = \, 1.39 + i(0.15 +
n)\,$.
\vskip0.5cm
\noindent
{\bf Fig}. 3. The dependence of $V $ on the $U_3 $ modulus for the
orbifold  $Z_6-II-b $. Minima occur at the fixed points of the
subgroup $\Gamma^{0}(3) \, $ acting on the redefined field
$U_3 - 2i $,  $ U_3 \, = \, \sqrt{3} /2 + i\,(3.5
+ 3 p) $.

\vfill\eject
\centerline {TABLE 1}
\vskip 0.5cm
\begintable
Orbifold |Point group generator|Lattice\cr
$Z_4-a$ | $(1,1,-2)/4$|$SU(4)\times SU(4)$
\cr
$Z_4-b$ | $(1,1,-2)/4$|$SU(4)\times SO(5)\times
SU(2)$
\cr
$Z_6-II-a$ | $(2,1,-3)/6$|$SU(6)\times
SU(2)$ \cr
$Z_6-II-b$ | $(2,1,-3)/6$|$SU(3)\times SO(8)$ \cr
$Z_6-II-c$ | $(2,1,-3)/6$|$SU(3)\times SO(7)\times
SU(2)$ \cr
$Z_8-II-a$ | $(1,3,-4)/8$|$SU(2)\times SO(10)$ \cr
$Z_{12}-I-a$ | $(1,-5,4)/12$|$E_6$
\endtable
\vskip4cm
\centerline {TABLE 2}
\vskip 0.5cm
\begintable
Orbifold | $C_{im} $|$l_{im} $\cr
$Z_4-a$ | $C_{31} = 2, C_{41} = 2 $|
$ l_{31} = 2, l_{41} = 1 $ \cr
$Z_4-b$ | $C_{31} =   C_{32} = C_{41} = C_{42} = 1$ |
$l_{31} = l_{41} = 1, l_{32} = 2, l_{42} = 1/2 $\cr
$Z_6-II-a$ | $C_{11} = 2, C_{31} = C_{32} = C_{41} = C_{42} =1$ |
$l_{11} = 2, l_{31} = l_{41} = 1, l_{32} = 3, l_{42}= 1/3$ \cr
$Z_6-II-b$ | $C_{11} = 2, C_{31} = C_{32} = C_{41} = C_{42} = 1$|
$ l_{11} =  l_{31} = l_{42} = 1, l_{32} = l_{41} = 3$ \cr
$Z_6-II-c$ | $C_{11} = 2, C_{31} = C_{32} = C_{41} = C_{42} = 1$ |
$l_{11} = l_{31} =l_{42} = 1, l_{32} = 3, l_{42} = 1/3$ \cr
$Z_8-II-a$ | $C_{31} = C_{32} = C_{41} = C_{42} = 1$|
$l_{31} = l_{41} = 1, l_{32} = 2, l_{42} = 1/2 $ \cr
$Z_{12}-I-a$ | $C_{31} = 2$|$l_{31} = 2 $
\endtable
\vskip 0.5cm
\epsffile[36 36 576 756]{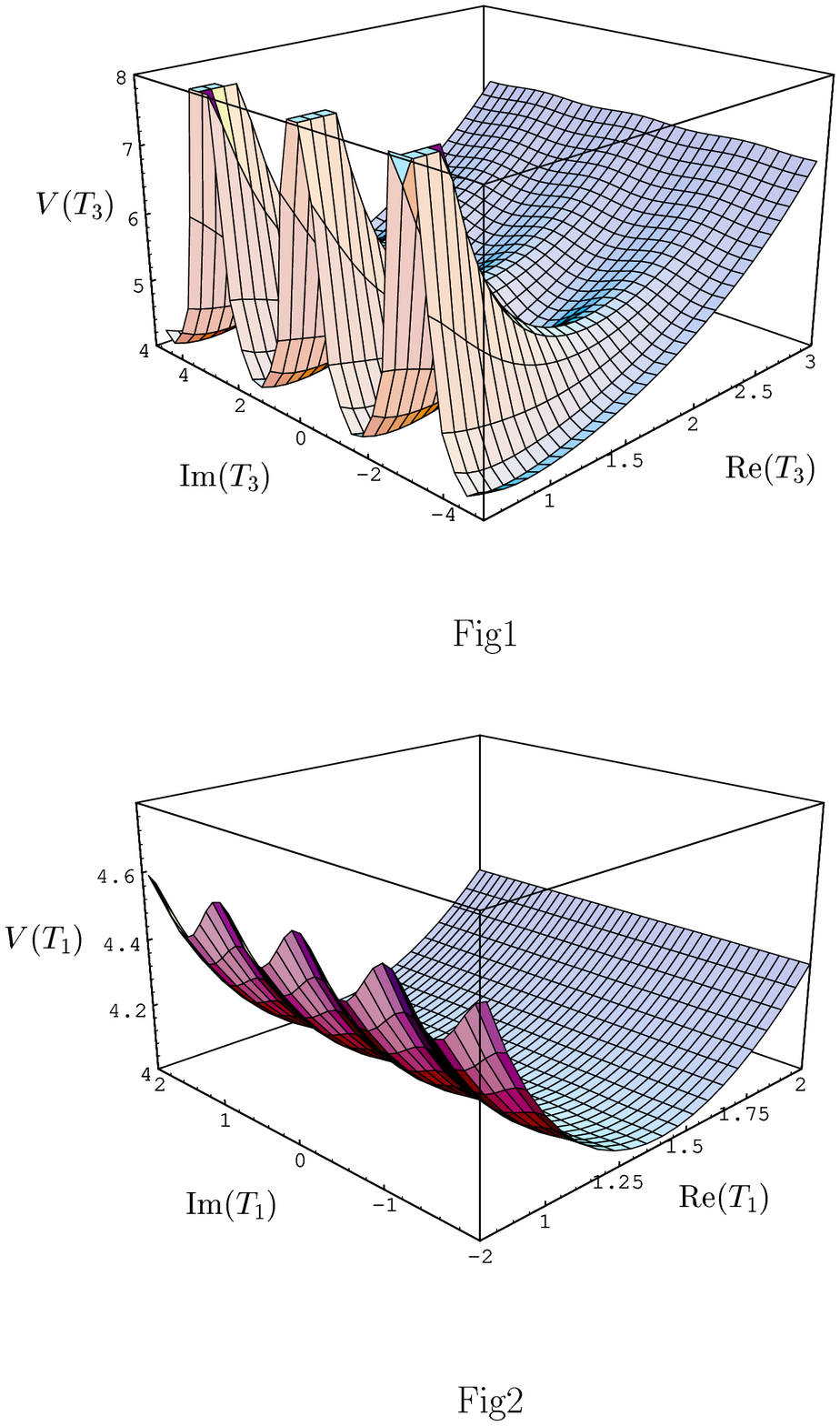}
\epsffile[36 36 576 756]{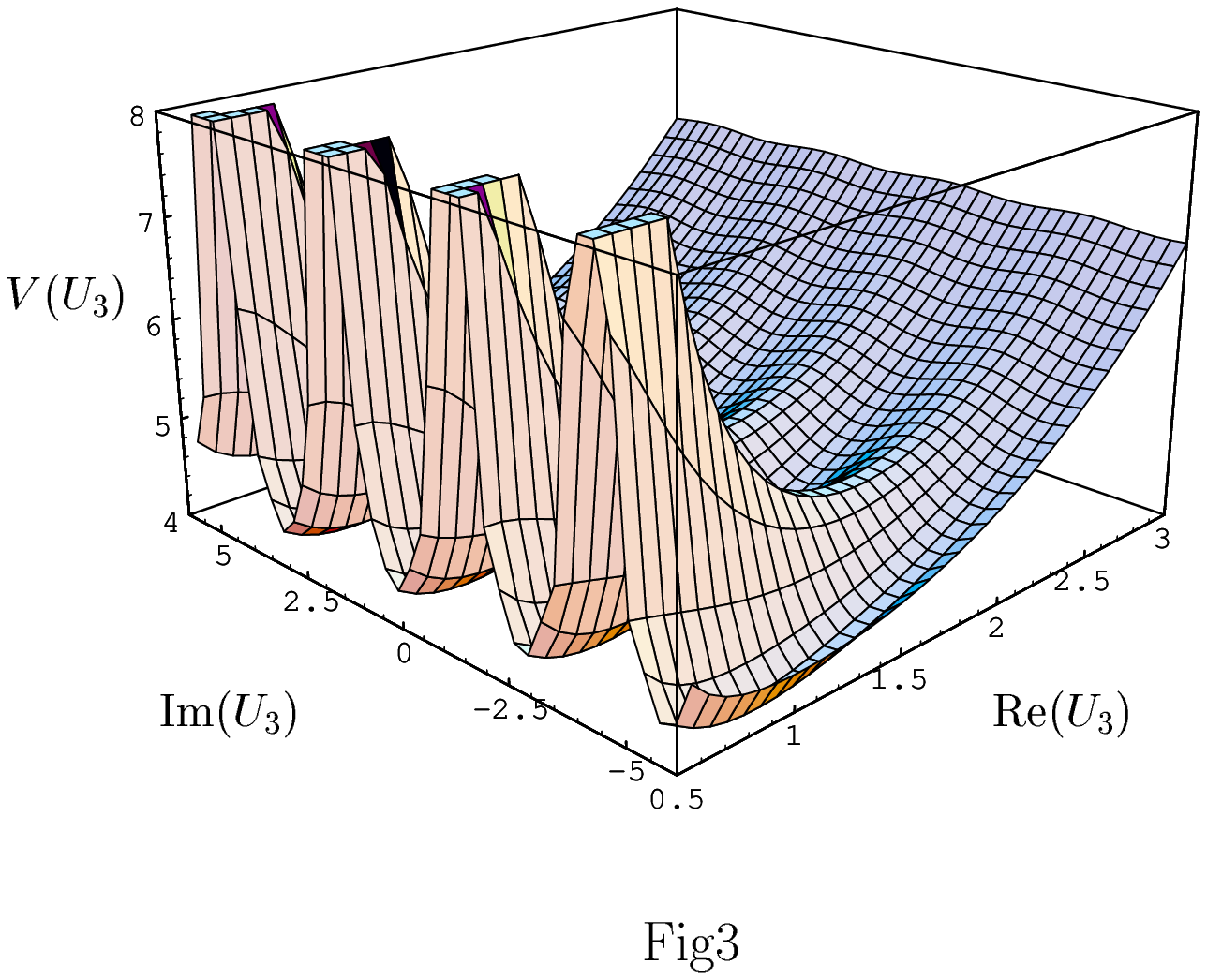}
\refout
\end